# Assessment of nanoparticle immersion depth at liquid interfaces from chemically equivalent macroscopic surfaces

**Joeri Smits[a], Rajendra Prasad Giri[b], Chen Shen[c], Diogo Mendonça[a,d], Bridget Murphy[b,e], Patrick Huber[c,f,g], Kurosch Rezwan[a,h], Michael Maas*[,a,h]**

## Author addresses

[a] Advanced Ceramics, University of Bremen, Am Biologischen Garten 2, D-28359 Bremen, Germany

[b] Institute of Experimental and Applied Physics, Kiel University, D-24098 Kiel, Germany

[c] Deutsches Elektronen-Synchrotron DESY, Notkestraße 85, D-22607 Hamburg, Germany

[d] Department of Mechanical Engineering, Federal University of Santa Catarina, 88040-900 Florianopolis, Brazil

[e] Ruprecht-Haensel Laboratory, Kiel University,24118 Kiel, Germany

[f] Hamburg University of Technology, Institute for Materials and X-Ray Physics, Eißendorfer Straße 42, 21073 Hamburg, Germany

[g] Hamburg University, Center for Hybrid Nanostructures ChyN, Luruper Chaussee 149, 22607 Hamburg, Germany

[h] MAPEX Center for Materials and Processes, University of Bremen, Bibliothekstraße 1, D-28359 Bremen, Germany

**E-mail addresses**

smits@uni-bremen.de (J. Smits), giri@physik.uni-kiel.de (R.P. Giri), chen.shen@desy.de (C. Shen), diogom96@hotmail.com (D. Mendonça), murphy@physik.uni-kiel.de (B. Murphy), patrick.huber@tuhh.de (P. Huber), krezwan@uni-bremen.de (K. Rezwan), michael.maas@uni-bremen.de (M. Maas*)

* Corresponding author

# Abstract

Hypothesis:

We test whether the wettability of nanoparticles (NPs) straddling at an air/water surface or oil/water interface can be extrapolated from sessile drop-derived macroscopic contact angles (mCAs) on planar substrates, assuming that both the nanoparticles and the macroscopic substrates are chemically equivalent and feature the same electrokinetic potential.

Experiments:

Pure silica ($SiO_2$) and amino-terminated silica ($APTES-SiO_2$) NPs are compared to macroscopic surfaces with extremely low roughness (root mean square [RMS] roughness $\leq 2$ nm) or a roughness determined by a close-packed layer of NPs (RMS roughness $\sim 35$ nm). Equivalence of the surface chemistry is assessed by comparing the electrokinetic potentials of the NPs via electrophoretic light scattering and of the macroscopic substrates via streaming current analysis. The wettability of the macroscopic substrates is obtained from advancing (ACAs) and receding contact angles (RCAs) and *in situ* synchrotron X-ray reflectivity (XRR) provided by the NP wettability at the liquid interfaces.

Findings:

Generally, the RCA on smooth surfaces provides a good estimate of NP wetting properties. However, mCAs alone cannot predict adsorption barriers that prevent NP segregation to the interface, as is the case with the pure $SiO_2$ nanoparticles. This strategy greatly facilitates assessing the wetting properties of NPs for applications such as emulsion formulation, flotation, or water remediation.

## Abbreviations

ACA, advancing contact angle; AFM, atomic force microscope; APTES, (3-aminopropyl)triethoxysilane; A/W, air/water; CA, contact angle; CAH, contact angle hysteresis; ECA, equilibrium contact angle; ED, electron density; ELS, electrophoretic light scattering; EOS, equation of state; FreSCa, freeze-fracture shadow-casting; GISAXS, grazing-incidence small-angle X-ray scattering; GTT, gel-trapping technique; mCA, macroscopic contact angle; NP, nanoparticle; OHW, Ohshima-Healy-White; O/W, oil/water; PF, particle-covered film; RCA, receding contact angle; RMS, root mean square; SC, streaming current; SEM, scanning electron microscope; SF, smooth film; SHG, second harmonic generation; $SiO_2$, silica; TEM, transmission electron microscope; TEOS, tetraethyl orthosilicate; XRR, X-ray reflectivity.

# 1. Introduction

Nanoparticles (NPs) adsorbing at a quiescent liquid interface are presented with a highly complex energetical and topographical landscape originating from contributions of the surface tension [1], capillary waves [2], viscoelasticity [3], and particle-particle interactions [4], giving rise to surface-specific behavior such as electrodipping forces [5], contact line undulations [6], and line tension contributions [7,8] that are dynamic in nature [9–11]. Particle wetting at the interfacial contact line, which defines interfacial adsorption, is conventionally approximated by macroscopic contact angle (mCA) measurements on a planar substrate with a surface composition and roughness that mirrors that of one adsorbed particle according to the Young equation [12]:

$$\cos \theta_Y = \frac{\gamma_{sg} - \gamma_{sl}}{\gamma_{lg}} \tag{1}$$

where Young's three-phase CA ($\theta_Y$) measured tangentially of the drop profile is determined by the interfacial tensions between each phase, namely solid-gas ($\gamma_{sg}$), solid-liquid ($\gamma_{sl}$), and liquid-gas ($\gamma_{lg}$), where the gas phase is exchangeable with an immiscible oil. Eq. 1 implies that wetting with a certain liquid provides a well-defined CA at thermodynamic equilibrium on a perfectly smooth surface, but this is rarely observed experimentally giving rise to contact angle hysteresis (CAH) from pinning/depinning phenomena [13,14]. Additionally, line tension effects ($\tau$) become important for nanoscopic drops or particles, which can be positive or negative in sign, and deviate from Young's CA [7,8]:

$$\cos \theta_i = \cos \theta_Y - \frac{\tau}{\gamma_{lg} b} \tag{2}$$

where $\theta_i$ is the line tension corrected CA and $b$ is the lateral radius of the drop or adsorbed particle. The line tension magnitude is estimated around $10^{-11}$ to $10^{-12}$ N but values up to $10^{-6}$ N have been measured and the inconsistency in sign and magnitude of $\tau$ is strongly debated [7,15,16]. In the case of a spherical particle with radius $a$ trapped at a liquid surface, a similar triple point is generated at the liquid-gas-solid intersection where the particle CA ($\theta_P$) is determined based on their water immersion depth ($h$) relative to the surface [17]:

$$\cos \theta_P = \left(\frac{h}{a}\right) - 1 \tag{3}$$

Eq. 3 does require exact knowledge about the position of adsorbed particles and becomes experimentally challenging when their size decreases. However, reported literature values of particle CAs and immersion depths show a distribution owing to dynamic adsorption and vary widely, in part stemming from the large mCA distributions measured between surfaces composed of the same material [9–11,18–20]. While alternative advanced methods were developed to measure CAs of adsorbed particles either directly (freeze-fracture shadow-casting (FreSCa) [20], gel-trapping technique (GTT) [21]) or indirectly (second harmonic generation (SHG) [19], ellipsometry [22], light extinction [23], X-ray reflectivity (XRR) [24,25]), accessibility to these fairly sophisticated methods is limited. Since interfacial adsorption of certain particle materials depends on pH [26] and ionic strength [26,27], which are directly related to the electrokinetic ($\zeta$) potential, efforts have been conducted to construct a wettability profile in relation to the $\zeta$ potential without determination of the mCA [28]. The exact immersion depth of adsorbed particles cannot be inferred from these studies, but if the $\zeta$ potential of particle dispersions determines their preferred wetting state, it also affects the measured mCA on a planar substrate that is a chemical and topographical equivalent of the adsorbed particles [29]. A direct association of $\zeta$ potential and CA values between particles and flat substrates could provide a platform with better reliability for assessing particle wettability.

Therefore, a versatile approach is tested (Fig. 1) where XRR-derived CAs of highly charged, hydrophilic silica ($SiO_2$) NPs or $SiO_2$ NPs coated with the aminosilane APTES (APTES-$SiO_2$) adsorbed to the air/water surface and decane/water interface are compared to mCAs of sessile droplets on corresponding planar substrates in an identical environment. The synthesis and functionalization procedures are adjusted to ensure that both surfaces have comparable surface chemistry, which is assumed to be the case when the $\zeta$ potentials of the dispersed NPs measured with electrophoretic light scattering and the coated planar substrates obtained through streaming current overlap. To understand the possible influence of surface roughness of the macroscopic substrate on the adsorption behavior of the particles, smooth (SFs) and particle-covered films (PFs) are evaluated concurrently.

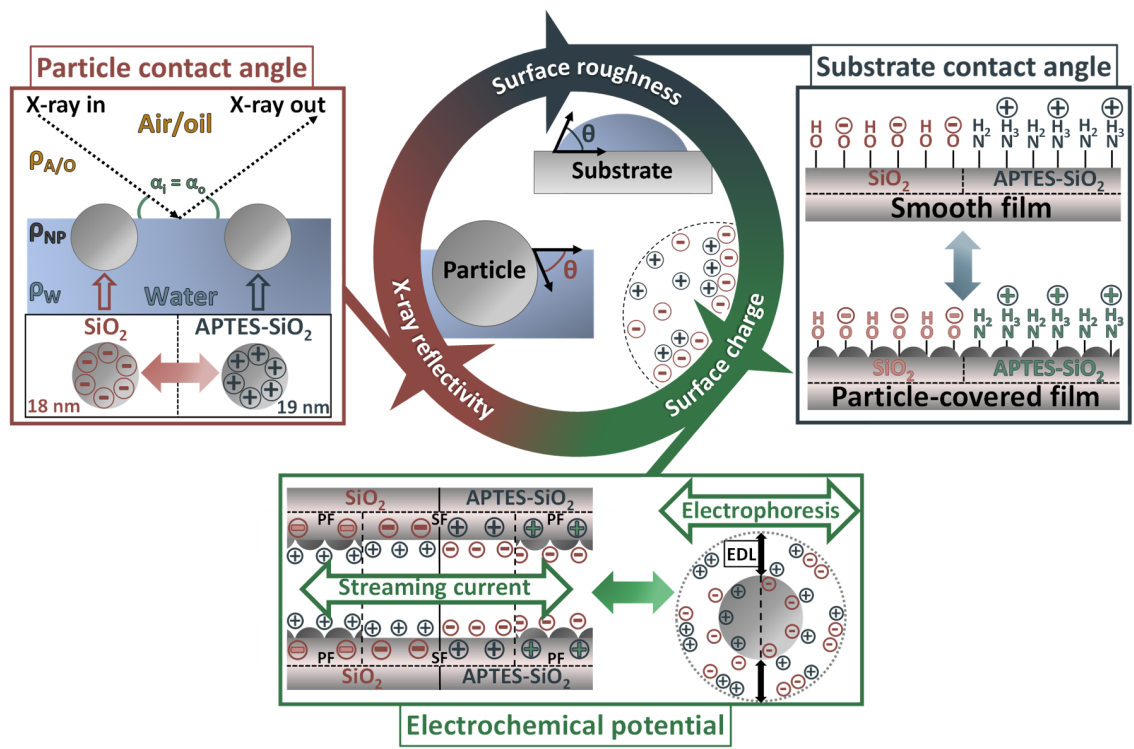

**Fig. 1 Overview of the methodology to relate the particle contact angle at liquid surfaces to macroscopic contact angles on planar substrates.** The particles and planar substrates consist of silica ($SiO_2$) or are coated with APTES (APTES-$SiO_2$). The liquid interface in presence of 18 nm $SiO_2$ and 19 nm APTES-$SiO_2$ is analyzed with synchrotron X-ray reflectometry from which the particle contact angle is derived. The zeta potential of particles and planar substrates is measured and compared using electrophoretic light scattering and streaming current, respectively. The effect of surface roughness is investigated by comparing the wettability of a smooth surface (smooth film, SF) with a particle-induced rough surface (particle-covered film, PF).

## 2. Materials and methods

### 2.1 Materials

Tetraethyl orthosilicate (TEOS $\geq$ 99.0%, product no. 86578), (3-aminopropyl)triethoxysilane (APTES 99.0%, product no. 440140), L-arginine ($\geq$ 99.0%, product no. 1.01542), potassium chloride powder (KCl 99.0 – 100.5%, product no. P3911) and liquid (MQ 100, 3 M KCl, product no. 60135), sulfuric acid ($H_2SO_4$ 95.0-98.0%, product no. 258105-M), Wash-N-Dry™ coverslip rack (product no. Z688568), Wheaton Coplin staining jars (product no. S6016), Hellma® optical glass cuvettes (volume 3.5 mL,45 x 12.5 x 12.5 mm, product no. Z600008) and tweezers (product no. T5540) were purchased from Sigma Aldrich (Darmstadt, Germany). Decane ($\geq$ 99.7%, product no. D901), HCl (0.1 M, product no. 38280; 1 M, product no. 71763), and NaOH (1 M, product no. 71463) solutions were purchased from Honeywell. Ethanol absolute (> 99.8%, product no. 64-17-5), glacial acetic acid ($\geq$ 99.7%, product no. 64-19-7), 37% HCl (product no. 20252.335), methanol (> 99.8%, product no. 20847.295), and amber Duran pressure plus bottles (product no. #215-3267) were obtained from VWR Chemicals (Hannover, Deutschland). Technical grade ethanol (99.0%, product no. 2211.5000) was bought from Chemsolute (Hamburg, Germany). Amorphous silica powders with a size of 80 nm (> 99.9%, SIO2P080-1KG) and 1000 nm (> 99.9%, SIO2P100-01-1KG) were obtained from Fiber Optic Center (New Bedford, USA). Kapton foil (200 x 304 x 0.05 mm, product no. 536-3952) was obtained from RS Components GmbH (Frankfurt am Main, Germany) and ZelluTrans dialysis tubes (MWCO 6-8 kDa, product no. E665.1) from Carl Roth GmbH (Karlsruhe, Germany). Minisart® syringe filter polyethersulfone (PES) with 0.1 μm pore size (product no. 16553----------K) were obtained from Sartorius Stedim Biotech (Germany, Göttingen). Thermo Scientific Menzel-Gläser coverslips (24 x 24 x 0.13-0.16 mm, product no. 10755315) were obtained from Fisher Scientific (Darmstadt, Germany). Microtip; Tapered, 6.5 mm ¼" for ultrasonication was bought from Terra Universal (Milan, Italy). All aqueous solutions were made with double deionized water (18.2 MΩ•cm at 25 °C) from a Synergy Water Purification system (Millipore Corp., Massachusetts).

## 2.2 Experimental section

Careful preparation and handling were ensured, so that all chemicals, particle dispersions, and macroscopic substrates were of the highest purity and behaved chemically inert towards the used laboratory materials to ensure that all reported measurements are only caused by the described particles or substrates [30,31]. In the case of the coated substrates, one side did not come into contact with another surface and was used for the related measurements. All glassware was cleaned by rinsing thrice with technical grade ethanol, acetone, water, and ethanol absolute followed by covering with a lint-free wipe and further drying in air. All aqueous solutions were degassed before use in Duran pressure plus bottles and sealed with a PTFE coated cap. An overview of the different silica materials and their application is provided in Table 1.

**Table 1** Overview of the used silica ($SiO_2$) and APTES-coated silica (APTES-$SiO_2$) materials and their main purpose. ACA, advancing contact angle; AFM, atomic force microscope; ELS, electrophoretic light scattering; mCA, macroscopic contact angle; PF, particle-covered; RCA, receding contact angle; RMS, root mean square; XRR, X-ray reflectivity.

| Material | Size | Main purpose |
|----------|------|-------------|
| $SiO_2$ | 17.8 nm | ELS: $\zeta$ potential; XRR: air/water and decane/water interfacial adsorption and CA |
| | 80 nm | ELS: $\zeta$ potential, dip coating, preparation of PF-substrates |
| | 1000 nm | ELS: $\zeta$ potential |
| | Planar (smooth and particle-covered) | AFM: RMS roughness; mCA: air/water and decane/water ACA/RCA; SC: $\zeta$ potential |

| APTES-SiO$_2$ | 18.9 nm | ELS: $\zeta$ potential; XRR: air/water and decane/water interfacial adsorption and CA |
|---|---|---|
| | Planar (smooth and particle-covered) | AFM: RMS roughness; mCA: air/water and decane/water ACA/RCA; SC: $\zeta$ potential |

## 2.3 Cleaning and activation of coverslips

During the whole procedure, coverslips were kept wet. To hydroxylate the coverslips before dip coating, separated coverslips (24 x 24 mm) were placed in a Coplin jar and filled with fresh 1:1 methanol:HCl and gently shaken for at least 30 min. After the first acid wash, the samples were rinsed with copious amounts of water and placed in a second Coplin jar filled with H$_2$SO$_4$ with gentle shaking for at least 30 min. The samples were rinsed with water, then with methanol, and stored for up to one week in a methanol-filled Coplin jar. Prior to the dip coating procedure, 2 glasses were rinsed with water and placed in a 500 mL Duran beaker filled with 100 mL of water, covered with a glass Petri dish, and placed in an oil bath at 140 °C for 45 min to cause gentle boiling [32]. Finally, the glasses were transferred to water at room temperature and dried under a stream of nitrogen before the dip coating procedure.

## 2.4 TEOS sol-gel medium preparation with and without dispersing 80 nm nanopowder

All solutions were prepared in glass vials. The volume ratios used were 1:2:2 of TEOS:ethanol:water to prepare a total of 40 mL. The acidic water was prepared by adding 1 mL of 37% HCl per 10 mL of water (pH <1). Ethanol absolute was then added during magnetic stirring (20 x 6 mm, 500 rpm) followed by dropwise addition of TEOS, and stirring was continued for 1 h. Three holes were pierced in the cap and the sol was placed for 5 h in a 70 °C oven. Afterwards, a new cap was placed on the vial and the sol-gel was subsequently stored overnight at 20 °C [33]. The next day, cleaned and activated coverslips were dip-coated in the sol-gel medium to obtain smooth thin films. For preparing particle-covered films, 1.33 mL of 3 M KCl was added to 36.667 mL of the TEOS sol-gel medium and magnetically stirred for 1 min. To this, 2 g of 80 nm nanopowder was gently added to the sol-gel medium during bath sonication to end up with a 5 w/v% dispersion. In total, the sol-gel dispersion was sonicated for 1 h and subsequently ultrasonicated (tip 1/4") for 5 min with ice cooling. The particle-sol-gel was left undisturbed at 20 °C for 30 min before the dip coating procedure.

## 2.5 Dip coating of activated coverslips in sol-gel medium with and without 80 nm nanopowder

The nitrogen-dried coverslips were dip coated in the acid-catalyzed TEOS sol-gel using a custom dip coating device. A 50 mL beaker (high form) was filled up to 30 mL with the sol-gel allowing to submerge 80 % of the coverslip and allowing the drying process to take place inside the beaker above the sol-gel surface. Dip coating parameters in the absence of nanopowder were 120 s residence time, 5 mm min$^{-1}$ withdrawal speed, and 10 min drying. When the nanopowder was added, the withdrawal speed was set at 50 mm min$^{-1}$ while keeping the other parameters identical since lower withdrawal speeds caused incomplete particle coverage. Substrates were then placed in Wash-N-Dry™ racks and sealed in a humidified atmosphere for 48 h to complete the condensation reaction [33]. Finally, the samples were heat-treated at 110 °C for 60 min, stored in an upright position separated from each other in ambient conditions, and protected from dust. The 80 nm nanopowder yielded the best coating results while the other particle sizes led to incomplete coverage for a variety of tested parameters and were not used in this work.

## 2.6 APTES coating on smooth (SF) and particle-covered (PF) silica films

Only smooth silica films (SF-SiO$_2$) with an advancing air/water contact angle between 5° - 12° were used for subsequent APTES functionalization and used immediately after preparation.

Briefly, a concentrated, pre-hydrolyzed alcoholic solution of APTES (1:0.95:0.05 ethanol absolute:APTES:water) was prepared one day in advance and stored at 4 °C [34]. A 50 mL glass vial with a plastic cap was filled with 15 mL of 10 v/v% ethanol absolute and 1 mM acetic acid, a hole was pierced in the cap, covered with alumina foil, and placed in a 90 °C water bath for 15 min [35]. The APTES stock solution was diluted 0.4:500 (0.04% APTES) in the heated solution, shaken, and SF- or PF-SiO$_2$ substrates were carefully immersed in the medium for 10 min at 90 °C. The samples were rinsed with 10 v/v% ethanol absolute, dried with nitrogen, and cured at 120 °C in the air for 30 min [36]. Samples were allowed to cool down for 15 min before dipping in 4:1 ethanol absolute:acetic acid, dried with nitrogen, and finally stored in 0.1 M HCl for 3 h [37]. Samples were rinsed with copious amounts of water and dried under a stream of nitrogen prior to the measurements.

## 2.7 Preparation of commercial nanopowder

The nanopowders (80 nm and 1000 nm silica) were subjected to 3 redispersion cycles in water. First, 5 g of powder was transferred into a 50 mL falcon tube, and water was added up to 45 mL. The falcon tubes were alternately vortexed and sonicated within 15 min followed by centrifugation of 2 or 10 min at 5000 rpm for 1000 nm and 80 nm silica, respectively. This step was repeated 3 times with fresh water. The particles were placed in a clean, glass Petri dish and covered with a lint-free wipe to air dry. After 72 h, the powders were collected in a glass vial and redispersed to 1 w/w% with alternating sonication/vortexing cycles and pH adjustments before characterization.

## 2.8 Synthesis of silica and APTES-coated silica nanoparticles

Silica nanoparticles were firstly synthesized [38] and additionally coated with APTES using a modified approach based on reference [39]. A concentrated, pre-hydrolyzed alcoholic solution of APTES (1:0.95:0.05 ethanol absolute:APTES:water) was prepared one day in advance and stored at 4 °C [34]. First, 300 mg L-arginine was dissolved in 300 mL water in a 500 mL round bottom flask and stirred magnetically (40 x 10 mm) with reflux at 70 °C for 1 h to ensure homogenization of the solution. Next, 32 mL of TEOS was added dropwise (20 mL h$^{-1}$) using a syringe pump with stirring at 1000 rpm up to 1 h after TEOS addition. The stirring was interrupted and TEOS was allowed to phase-separate for 15 min. A funnel of TEOS was created by setting the speed at 250 rpm for at least 24 h until all of the hydrophobic liquid was consumed. The produced silica nanoparticles were stored at 4 °C or coated with APTES. In the latter case, the oil bath was set to 40 °C and the particles cooled down in an ice bath to 4 - 6 °C. Acetic acid was diluted in ethanol absolute so that the nanoparticle dispersion contained 100 mM acetic acid and 10 v/v% ethanol, which was added dropwise at 40 mL h$^{-1}$ with high stirring and maintained 30 min post-addition. The dispersion was reheated at 40 °C with reflux while the pre-hydrolyzed APTES solution was allowed to thermally equilibrate to room temperature. At 10 mL h$^{-1}$, 20 mL of pre-hydrolyzed APTES was added to the dispersion (1000 rpm) and afterwards heated to 90 °C for 24 h (750 rpm). Another 100 mM acetic acid dissolved in 5 mL ethanol absolute was added dropwise until the final pH was approximately 5 and stored at 4 °C prior to purification.

## 2.9 Synthesized nanoparticle purification

APTES-coated silica particles were centrifuged for 30 min at 5000 rpm to sediment larger agglomerates while silica nanoparticles were filtered (PES 0.1 μm) before dialysis. About 30 mL of particles were transferred to a dialysis bag, placed in a 2 L beaker (high form) containing 1:0.5:0.007 of water:ethanol absolute:acetic acid with magnetic stirring for at least 4 h and repeated twice. The particles were then dialyzed 10 times against water (pH ~ 5.8) with a minimum time interval of 4 h. When dialysis was finished, the water was refreshed one final time and stirred for 30 min before collecting the particles in a clean Duran beaker enclosed with a PTFE-coated cap. The pH was adjusted to 5.8, the solid content was determined in triplicate to be $2.32 \pm 0.02$ for silica and $1.76 \pm 0.04$ w/w% for APTES-coated silica NPs and stored at 4 °C. The aqueous dispersions remained stable for at least 3 months based on dynamic light scattering.

## 2.10 Scanning Electron Microscopy (SEM), Transmission Electron Microscopy (TEM), and Atomic Force Microscopy (AFM)

The smooth (SF) and particle-covered (PF) films were visually examined with SEM (Supra 40, Carl Zeiss, Germany) as well as the 80 nm and 1000 nm silica nanoparticles. Transmission electron microscopy (TEM-EM 900, Carl Zeiss, Germany) to study the morphology of the smaller particles was done by placing a 3 μL drop of 0.001 w/w% particle dispersion (pH ~ 5.8) onto a copper grid (Plano GmbH, Wetzlar, Germany). AFM (JPK Nanowizard III, Berlin, Germany) on the substrates was performed using a cantilever TESPA-V2 from Bruker with tip radius 7 nm, nominal frequency of 320 kHz, and nominal spring constant of 37 N m$^{-1}$ in the tapping mode. The line rate was 1 Hz for smooth SF samples and 0.5 Hz for rough PF samples. Gwyddion software was used to determine the root mean square (RMS) roughness and averaged over 4 samples.

## 2.11 Air/water and oil/water contact angle measurements

### 2.11a Experimental

Contact angle (CA) goniometry of sessile water droplets measured on planar silica and APTES-coated silica substrates formed from a dosing needle (0.55 mm diameter) in air and immersed in decane was done using the OCA20 from Dataphysics (Stuttgart, Germany) at 20 °C (± 1 °C) with a standard error of 0.05°. All samples were rinsed with water and dried with nitrogen before CA measurements. Advancing (ACAs) and receding contact angles (RCAs) were measured with the needle-in sessile drop method according to the procedure proposed by Korhonen et al. [40] by simultaneously evaluating CA, drop volume, and base diameter. In general, a 1 µL water drop was formed on the substrate with a 10 s waiting time followed by slowly increasing the droplet volume until the monitored base diameter was at least 5 mm after which the water was retracted. Air/water and oil/water ACAs/RCAs on all samples were obtained at a dosing speed of 0.025 µL s$^{-1}$, except for SF-APTES where the speed was set at 0.05 µL s$^{-1}$. Samples for air/water CA measurements were prewetted by dropping a larger volume on the surface to bypass stick-slip behavior [41], which was retracted and discarded before placing a fresh 1 µL drop and measuring the ACA/RCA. For oil/water CA measurements, after cutting the samples using a diamond pen (10 x 10 mm), they were rinsed with water and dried with nitrogen before carefully placing them in a cuvette. The cuvette was quickly filled with 900 µL pure decane (test for purity see Supplementary Section S3, Fig. S3) and left to equilibrate for 30 min before obtaining the ACA/RCA (Supplementary Section S5, Fig. S5). It was not possible to measure the RCA on all substrates due to the vanishing drop contour and a total of 50 samples ($n = 50$) were measured for each type of surface (Supplementary Section S6, Table S6).

**2.11b Estimating wettability parameters**

Equilibrium contact angles (ECAs, $\theta_E$), which from a thermodynamic point of view is the lowest energy shape for a drop on a surface, have been calculated from ACAs ($\theta_A$) and RCAs ($\theta_R$) using a model that implements the contact line energy into the Young-Laplace equation [42]:

$$\theta_E = \cos^{-1}\left(\frac{\Gamma_A \cos\theta_A + \Gamma_R \cos\theta_R}{\Gamma_A + \Gamma_R}\right) \qquad (4)$$

where

$$\Gamma_A = \left(\frac{\sin^3\theta_A}{2\text{-}3\cos\theta_A + \cos^3\theta_A}\right)^{\frac{1}{3}} \tag{5}$$

$$\Gamma_R = \left(\frac{\sin^3\theta_R}{2\text{-}3\cos\theta_R + \cos^3\theta_R}\right)^{\frac{1}{3}} \tag{6}$$

The goniometer cannot measure CAs $\leq 5.00°$ accurately and when no RCA could be measured due to vanishing drop contour and significant spreading of the liquid (i.e. visually zero contact angle), a value of $0.10°$ or $5.00°$ was chosen for these calculations. The difference between $\theta_A$ and $\theta_R$ is the contact angle hysteresis ($\Delta\theta$) and can be expressed as a percentage to describe the degree of hysteresis, where 0% means no hysteresis and 100% maximum hysteresis [40]:

$$\Delta\theta_\% = \left(\frac{\cos\theta_R - \cos\theta_A}{2}\right)100 \tag{7}$$

$\Delta\theta_\%$ is linearly proportional to the work of adhesion, allowing to directly compare hysteresis values with each other how the movement of the contact line is facilitated on a specific surface. Chibowski [13,43] proposed to determine the total apparent surface free energy of a smooth, solid material ($\gamma_{sg}$) based on the contact angle hysteresis (CAH) with one test liquid in air:

$$\gamma_{sg}^{CAH} = \frac{\gamma_{lg}(1 + \cos\theta_A)^2}{2 + \cos\theta_R + \cos\theta_A} \tag{8}$$

The value for the air/water surface tension ($\gamma_{lg}$) is 0.0722 N m$^{-1}$. Additionally, $\gamma_{sg}$ can be estimated from Neumann's equation of state (EOS) approach from the ACA or the ECA determined from Eq. 4 [41]:

$$\cos\theta_Y = -1 + 2\sqrt{\frac{\gamma_{sg}^{EOS}}{\gamma_{lg}}}e^{-\beta\left(\gamma_{lg} - \gamma_{sg}^{EOS}\right)^2} \tag{9}$$

where the constant $\beta = 1.25 \times 10^{-4}$ $(m^2\ mJ^{-1})^2$. $\gamma_{sg}$ determined from Eq. 8 using ACA/RCA and Eq. 9 using the ACA gave similar results and their average was reported ($\gamma_{sg}^{Av}$). Eq. 9 is obtained by combining Eq. 1 with the EOS for the interfacial tensions to calculate the solid-liquid interfacial tension $\gamma_{sl}$:

$$\gamma_{sl}=\gamma_{lg}+\gamma_{sg}-2\sqrt{\gamma_{lg}\gamma_{sg}}e^{-\beta\left(\gamma_{lg}-\gamma_{sg}\right)^2} \tag{10}$$

To determine the solid-liquid-liquid surface free energy (i.e. at oil/water), Stammitti-Scarpone et al.[44] obtained reasonable predictions for the solid-oil interfacial tension ($\gamma_{so}$) using an extended-EOS (e-EOS) for aqueous droplets on a sample immersed in an oil:

$$\gamma_{so}=\gamma_{lo}+\gamma_{sl}-2\sqrt{\gamma_{lo}\gamma_{sl}}e^{-\beta\left(\gamma_{lo}-\gamma_{sl}\right)^2} \tag{11}$$

where $\gamma_{lo}$ is the oil/water interfacial tension (decane/water = 0.0522 N $m^{-1}$). Modifying the Young equation (Eq. 1) allows to predict the CA for a solid in presence of an immiscible oil and water ($\theta_{lo}$) from the ACA/RCA or the ECA:

$$\cos\theta_{lo}=\frac{\gamma_{so}-\gamma_{sl}}{\gamma_{lo}} \tag{12}$$

## 2.12 Electrophoretic light scattering for colloidal nanoparticles

To ensure all organic impurities were removed, 20 g of KCl was dissolved in 200 mL water and heated at 550 °C for 24 h. This purified salt was used in the preparation of the electrolyte solutions for both electrophoretic light scattering and streaming current measurements. Zeta potentials at different pH levels were obtained by preparing 0.1 w/w% particle dispersions in 10 mM KCl for conductivity at 25 °C and measured using the M3-PALS technology available for the ZetaSizer

NanoZSP (Malvern, United Kingdom). Adjustment of pH was done using a maximum of 3 µL doses of 0.1 M or 1 M KOH or HCl. A stock solution of 1 w/w% in pure water was first prepared for the nanopowders and sonicated for 30 min before diluting to 0.1 w/w%. The nanopowder was allowed to equilibrate (rehydroxylation) until the pH was stable, creating a nano-environment that is more representative of non-dried particles [45]. The pH of the purified nanopowder was adjusted twice per day with magnetic stirring in between until the pH reading after equilibration time was 0.3 units away from the desired pH. At this point, the pH was slightly adjusted to the desired value and stirred for 1 min before measuring. Conversely, fewer pH fluctuations were observed in the case of the synthesized particles. These particles were magnetically stirred for 15 min after pH adjustment followed by a final pH reading right before electrophoretic light scattering. A waiting time of 300 s was used to ensure temperature calibration of the samples with an average of 3 samples for each pH value. Electrophoretic mobilities were used to calculate the zeta potentials which depend on the Debye parameter ($\kappa$) and particle radius ($a$) (Supplementary Section S2a). Hydrodynamic diameters were obtained at a particle concentration of 0.1 w/w% at 25 °C with 3 averaged measurements, containing 10-13 sub-measurements with a duration of 10 s per iteration (Supplementary Section S2a, Table S2).

### 2.13 Streaming current for rectangular planar substrates

Zeta potentials of SF- and PF-substrates were obtained via the streaming current method using an electrokinetic analyzer SurPASS 2 (Anton Paar Germany GmbH) at 20 °C (± 1 °C) with an adjustable gap cell for rectangular samples. Before each measurement, the conductivity and pH electrodes were placed in each test liquid for a minimum of 15 min to avoid drift during calibration. Dry, planar substrates were precisely cut (10 x 20 mm) in a homogenous area of the sample, excluding the irregular edges. These slides were attached to the sample holders using double-sided adhesive tape and placed inside the cell. 1 L of fresh 10 mM KCl solution was prepared from degassed water and 100 mL of the solution was transferred to a 600 mL beaker with a stirrer bar (40 x 10 mm) for rinsing and subsequently discarded. The beaker was filled with 500 mL electrolyte, pH, and conductivity electrode rinsed with the remaining electrolyte and the sample cell was rinsed for 20 min at a pressure of 200 mbar while adjusting the gap distance, which was generally around 110 µm. After 20 min, a flow check was performed (corresponding to a flow of 50 - 150 mL min$^{-1}$ at 200 - 400 mbar) and the 0.05 M KOH titration solution was rinsed to remove air bubbles. Titration was performed from pH ~ 5.8 (due to dissolved $CO_2$) to basic pH 9 at a pressure of 400 mbar and repeated in triplicate. Since measuring from this starting pH to basic has no influence on the zeta potential measurements, $N_2$ purging as suggested by [46] was not implemented in our laboratory setup.

## 2.14 X-ray reflectivity

### 2.14a Experimental

The synchrotron reflectivity measurements (XRR) at the decane-water interface were performed at the Liquid Interface Scattering Apparatus (LISA) diffractometer at the P08 beamline, PETRA III for its delivery of X-rays with high brilliance (DESY, Hamburg, Germany) [47,48] and are described in detail in our recent publication [49]. Briefly, a 25 keV collimated X-ray beam with a size of 0.05 x 0.4 mm (vertical x horizontal direction) probes the gas/liquid surface and liquid/fluid interface while varying the incident angles ($\alpha_i$), without moving the sample, to obtain a reflectivity profile as a function of the momentum transfer $q = 4\pi sin(\alpha_i)/\lambda$, where $\lambda = 0.496$ Å. A custom-made cell hosting a Delrin cup (inner diameter 60 mm) was filled with 0.1 w/w% particle dispersion (pH ~ 5.8) to pin the meniscus and positioned perfectly horizontal on the goniometer. For the decane/water measurements, decane was slowly added from the top by a syringe to a thickness of about 1 cm to avoid decane/air scattering. Reflectivity measurements were initiated after 30 min equilibration time once the liquid surface/interface was formed. The background-corrected XRR data collected with a Lambda X-ray detector was analyzed using REFLEX software based on Parratt's recursive algorithm with known material electron density (ED) to obtain values for the layer thickness and interfacial roughness [50]. These parameters were highly constrained using literature values and prior knowledge to be kept within physical boundaries [51].

## 2.14b Physical model for interfacially adsorbed nanoparticles

The physical model used here was adopted from Isa et al. [24] A Fourier transform is utilized for fitting the experimental reflectivity data into an averaged ED profile $\varrho(z)$ within the plane of the reflecting interface to the surface normal. It is subsequently fitted to a model that takes the physical parameters of the interface and the shape of the NPs into account:

$$\varrho(z) = \varrho_i(z) + \iint \left[ \varrho_{NP}\left(s_{av}, \varrho_p; z\right) - \varrho_{NP}\left(s_{av}, \varrho_i(z); z\right) \right] G(a_{av}, \sigma_a; a) G(h_{av}, \sigma_h; h) da\, dh \qquad (13)$$

$\varrho_i(z)$ describes the bare decane-water interface with bulk electron densities (EDs) of water $\varrho_w = 0.334$ e Å$^{-3}$ and decane $\varrho_d = 0.253$ e Å$^{-3}$, and an interfacial roughness $\sigma_i$ described by a Gaussian error function:

$$\varrho_i(z) = \varrho_w + \frac{\varrho_d - \varrho_w}{2} \left[ 1 + \text{erf}\left(\frac{z}{\sigma_i\sqrt{2}}\right) \right] \qquad (14)$$

$\varrho_{NP}(z)$ represents the contribution of the NP adsorbed at the interface, modeled as a monolayer of hexagonally packed spheres of radius $a$, immersion depth $h$, average interparticle spacing $s_{av}$, and bulk particle electron density $\varrho_p$:

$$\varrho_{NP}(z) = \varrho_{NP}\left(s_{av}, \varrho_p; z\right) \tag{15}$$

$$= \frac{2\pi\varrho_p}{\sqrt{3}(2a + s_{av})^2} \left[-z^2 + 2(a-h)z + h(2a-h)\right] \tag{16}$$

when $-h < z < -h + 2a$ and $\varrho_{NP}(z) = 0$ otherwise. Liquid displacement due to NP adsorption needs to be accounted for, which represents the subtracted term in Eq. 13. Normalized Gaussian distributions $G(x_{av}, \sigma_x; x)$ (Eq. 13) are introduced to account for size polydispersity (*a*) and consequently varying immersion depths (*h*) with their average values $x_{av}$ and their widths $\sigma_x$. Particle size, electron density, and interfacial roughness were fixed to their known values or constrained and the average separation distance between the NPs ($s_{av}$) and the average immersion depth ($h_{av}$) was allowed to vary. The former determines the average particle coverage associated with the height of $\varrho(z)$, while the latter describes the peak particle contact angle by employing Eq. 3 and the z-coordinate of the maximum in the ED profiles.

## 3. Results and discussion

### 3.1 Mimicking the nanoparticle surface on smooth and rough planar substrates

Comparable nano- and macroscale surfaces were prepared via the sol-gel strategy using the alkoxides tetraethyl orthosilicate (TEOS) and (3-aminopropyl)triethoxysilane (APTES). Stöber-silica ($SiO_2$) particles were prepared with TEOS at moderate basic pH, while the same alkoxide used at acidic conditions forms a weakly-branched polymeric network and is transferred to a glass substrate by dip coating [52]. Subsequent drying results in coated substrates with smooth surface films (SF). While different degrees of roughness largely influence the immersion depth of adsorbed microparticles [53], increased roughness on planar substrates may better represent the wetting behavior of adsorbed NPs compared to smooth coatings. Particle-covered films (PFs) with increased roughness were prepared by adding 80 nm silica particles to the sol-gel medium before the coating (SEM image shown in Supplementary Section S1, Fig. S1). The particles are trapped on the surface and are covered by a thin silane film upon drying, forming a stable surface that can withstand liquid shear forces (e.g. during streaming current experiments). In a second step, APTES was covalently attached to the TEOS surface of both NPs and substrates by incubating the materials in the respective sol-gel medium. The experimental procedure described in the Methods section should ensure that multilayer formation of APTES is suppressed due to the short reaction time of pre-hydrolyzed APTES in a diluted alcoholic solvent at elevated temperatures followed by rinsing with acetic acid [34,35,54,55]. The topographical surface roughness of the dip-coated planar substrates is evaluated using AFM and SEM as shown in Fig. 2. The average root mean square (RMS) roughness of the substrate before APTES-coating (SF-$SiO_2$) is 0.358 nm and very close to that of mica (0.2 nm) [56]. The average RMS roughness increases to 2.178 nm after APTES functionalization (SF-APTES-$SiO_2$). This increased roughness is likely a combination of silica-etching [57] with possible swelling of APTES at the elevated processing temperature of 90 °C [58] and low APTES concentration of 0.04% leading to incomplete coverage [59]. The presence of 80 nm silica NPs confined at the surface before (PF-SiO$_2$) and after APTES-functionalization (PF-APTES-SiO$_2$) can be observed with both AFM and SEM in Fig. 2. Although not shown, some micron-sized particle aggregates are present on the PF substrates owing to their high stability in the sol-gel medium, which leads to irregular deposition of the NPs. Subsequent APTES functionalization of PF-SiO$_2$ shows no difference in RMS roughness, which lies around 35 nm.

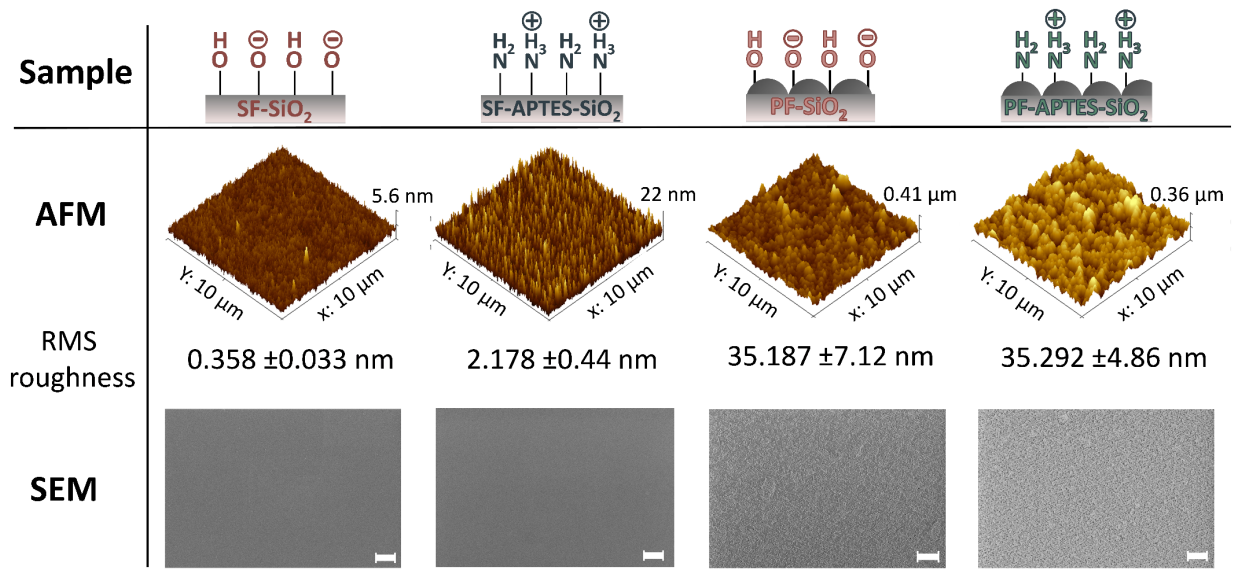

**Fig. 2 Topographical analysis of planar smooth (SF) and particle-covered films (PF) of silica (SiO$_2$) and APTES-coated silica (APTES-SiO$_2$) with AFM and SEM.** A 3D surface area of 10 x 10 µm obtained with AFM is shown from which the average root mean square (RMS) roughness per type is determined with standard deviation ($n$ = 4). The bottom row depicts SEM images with scale bars of 1 µm.

## 3.2 Analogous electrokinetic potentials between nanoparticles and macroscopic substrates

The reported correlation of particle wettability and electrokinetic ($\zeta$) potential provides a basis to estimate wettability at both nano- and macroscale [29]. At the nanoscale, the $\zeta$ potential is measured for differently sized colloids with electrophoretic light scattering, where charged particles migrate to the counter electrode when an external electric field is applied (Supplementary Section S2a1).

For the macroscopic planar substrates, an electric current is evoked when a salt solution is sheared over a stationary surface. The streaming current arises due to charge transfer when the sheared counterions reach the electrode (Supplementary Section S2b).

However, comparative $\zeta$ potentials between both methods necessitate certain conditions. Briefly, determination of $\zeta$ potentials for smooth and spherical particles based on Smoluchowski's (Eq. S1), Hückel's (Eq. S3) and Henry's (Eq. S4) model show a linearly varying electrophoretic mobility ($\mu$) with respect to $\zeta$ potential and $\kappa a$ ($\kappa$ is the Debye parameter (Eq. S2) describing the electric double layer thickness $\kappa^{-1}$, which is inversely related to the electrolyte concentration and $a$ the particle radius). Though, the diffuse double layer around dispersed colloids attains an asymmetric shape for highly charged colloids ($\zeta \geq 25$ mV and $\kappa a \geq 2.75$) that generally underestimates $\mu$ and becomes more pronounced with smaller $\kappa a$ such as with decreasing particle size and/or less added electrolyte [60]. Based on the numerical results from O'Brien and White [60], approximate analytical expressions have been designed to account for this relaxation or double-layer polarization which describes a non-monotonic relationship between $\mu$ and $\zeta$ potential. Approximate solutions, such as those by Ohshima, Healy, and White (OHW, Eq. S8) [61] and by Ohshima (Eq. S9) [62], cover the whole range of possible $\kappa a$ values. Determination of $\kappa a$ for the colloidal silica particles depicted in Fig. 3a is discussed in detail in Supplementary Section S2a2 and model-dependent $\zeta$ potentials are visually shown in Fig. S2b. Importantly, at $\zeta$ potentials < 50 mV and electrolyte concentrations $\geq 10$ mM, the effects of surface conduction can often be ignored, which would otherwise lower experimental $\mu$ due to increased counterion adsorption compared to the bulk by an excess of electrical charge on the solid surface. However, surface conduction becomes increasingly problematic when the particle size decreases and since it is not accounted for in Ohshima's approximation ($\kappa a < 10$), it could underestimate $\zeta$ potentials even below 50 mV [63–65]. This issue is tackled by using very large particles (1000 nm $SiO_2$) where surface conduction is practically absent, directly determined from Dukhin's number ($\leq 0.05$) in the OHW equation, and assuming that their surface chemistry is equal to the smaller sized silica particles [45]. Fig. 3a shows the $\zeta$ potentials obtained through electrophoresis for the 3 differently sized silica colloids at a fixed particle concentration of 0.1 w/w% dispersed with a background electrolyte concentration of 10 mM KCl. The $\zeta$ potentials are almost identical when pH $\leq$ 8, which can be expected for chemically indistinguishable surfaces. Therefore, micrometer-sized spheres can be used as a reference to determine the $\zeta$ potential in relation to corresponding particles of smaller sizes. For completeness, the $\zeta$ potential values of the 19 nm APTES-coated silica NPs were also calculated with Ohshima's expression [62].

The streaming currents of the smooth (SF) and rough (PF) planar substrates composed of silica ($SiO_2$) or coated afterward with APTES (APTES-$SiO_2$) in Fig. 3b are calculated using the Helmholtz-Smoluchowski equation (Supplementary Section S2b, Eq. S10) since $\kappa a \gg 1$ and surface conductance above an electrolyte concentration of 1 mM is found to be insignificant for silica during streaming current experiments [66,67]. PFs for both surface chemistries show slightly weaker $\zeta$ potentials compared to the SFs. The AFM images for the PFs in Fig. 2 show a varied landscape with peaks and valleys, thus it is very likely that there is no well-defined shear plane within the double layer as presumed by the Helmholtz-Smoluchowski model. The mobile ions inside the valleys compared to the peaks remain mostly stationary during shearing and therefore only part of the surface is being detected at the electrodes [67].

Finally, the $\zeta$ potentials of nanoparticles and macroscopic substrates overlap after proper conversion and determination of $\zeta$ potentials for dispersed particles with electrophoretic light scattering and thin films with streaming current (Fig. 3c, excluded PF data). It should be noted that, although size-dependent $\zeta$ potential is accounted for by considering $\kappa a$, the surface charge density ($\sigma$) will increase significantly when the particle size decreases below $a = 5$ nm in comparison to a non-porous, flat surface [68,69]. However, this effect is still moderate for the smallest particle size used (Supplementary Section S2c, Fig. S2c). In this case, the direct connection between $\sigma$ and wettability, both at liquid surfaces [70] and on planar substrates [71], could lead to different CAs when working with ultrasmall particles which pose a limit in relating the mCA to the immersion depth of NPs at liquid interfaces [29].

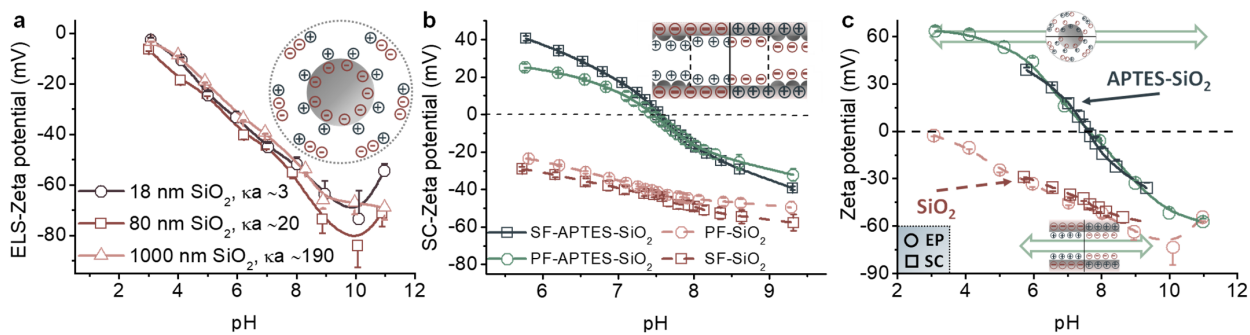

**Fig. 3 Evaluation of zeta potential with electrophoresis and streaming current at 10 mM KCl.** **a**, Zeta potential measurements using electrophoretic light scattering (ELS) for 0.1 w/w% of 1000, 80, and 18 nm silica ($SiO_2$) particles at different pH with their approximate dimensionless electrokinetic radius $\kappa a$. **b**, Zeta potential using the streaming current (SC) method for smooth (SF) and particle-covered films (PF) with $SiO_2$ or coated with APTES (APTES-$SiO_2$) are measured in a pH range from 5.8 to 9.4. **c**, Comparing the zeta potentials obtained with the ELS and SC method of $SiO_2$ and APTES-$SiO_2$. Only results of the SF-samples from **3b** are depicted in **c**. The maximum range of the pH titration was different for both methods and is indicated schematically by the broad arrows and experimental standard deviations are provided in all plots ($n = 3$).

### 3.3 *In-situ* observation of self-assembled nanoparticles at liquid interfaces

The NPs self-assemble at a quiescent air/water surface and oil/water interface, with an immersion depth that can be determined *in situ* from high-resolution X-ray reflectivity (XRR) data. This discussion is based on a dataset and analytical approach that was also used in our recent publication which focuses on an in-depth discussion of particle assembly at the oil/water interface in presence of oil-soluble surfactants [49]. The purity of the particle dispersions (0.1 w/w% at aqueous pH ~ 5.8) and decane are tested with time-resolved interfacial tension measurements before the XRR tests (Supplementary Section S3, Fig. S3). The Fresnel normalized reflectivity profiles ($R/R_F$) in Fig. 4 indicate that when the subphase contains negatively charged silica NPs, adsorption is mostly absent since the reflectivity curve resembles that of a pure air/water surface with a very small negative slope due to the interface roughness [72]. When particles are positively charged, pronounced Kiessig fringes appear at both liquid interfaces with an estimated thickness ($d \approx 2\pi/\Delta Q_z$) of 18.74 ± 1.45 nm, similar to the particle diameters estimated from TEM (Supplementary Section S1, Fig. S1). The preference of the positively charged particles to adsorb at the interface can be linked to the negative charge of a pristine air/water surface or immiscible oil/water interface, while the negatively charged particles are electrostatically repelled [25,73,74], and to their inherent surface wetting properties [19]. The raw data is fitted with electron density (ED) profiles using the REFLEX software [50] to simulate the XRR data based on Parratt's recursive formalism [75] using a 3-slab model along with the interface normal and the corresponding EDs are fitted against a physically realistic model to obtain adsorption parameters (Supplementary Section S4, Table S4a) [49]. The EDs normalized against water ($\rho_w$) in the insets of Fig. 4 indicate that the APTES-coated silica NPs adsorb to both liquid surfaces with a small protrusion distance into the non-polar gas or liquid. The CA changes from 28.35° ± 2.52 to 22.26° ± 2.68 when air is replaced by decane, while all other parameters remain constant (Supplementary Section S4, Table S4b).

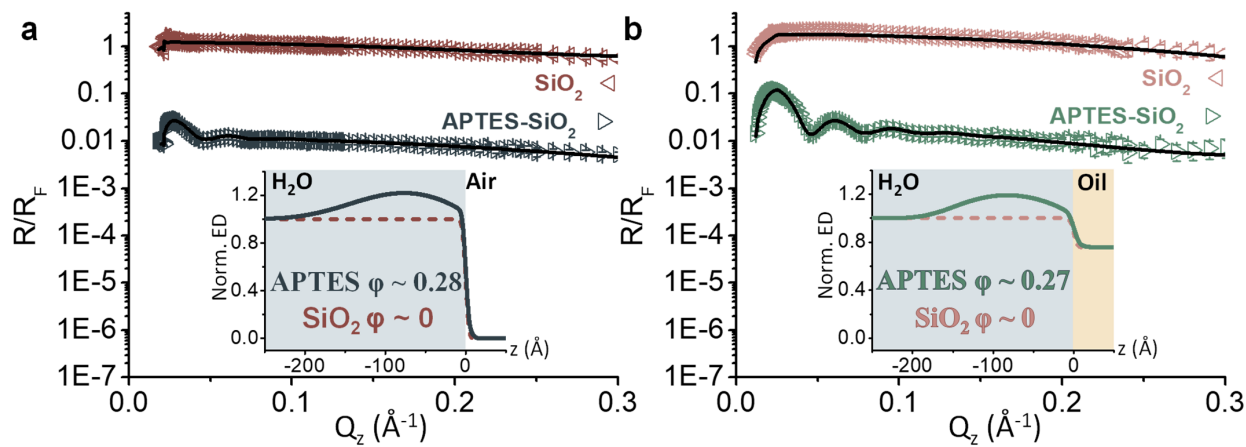

**Fig. 4 X-ray reflectivity and electron densities of adsorbed particles at the air/water surface and decane/water interface. a**, Reflectivity curves normalized to the Fresnel reflectivity ($R_F$) at the air/water interface and **b**, the decane/water interface (aqueous subphase pH ~ 5.8) with 0.1 w/w% 18 nm silica ($SiO_2$, left triangles) and 19 nm APTES-coated silica (APTES-$SiO_2$, right triangles) with their corresponding fits (black line) and standard deviation ($n = 3$). The insets show the normalized-to-water ($\rho_w$) electron density plots for $SiO_2$ (dashed line) and APTES-$SiO_2$ (solid line) from which the physical parameters, such as particle coverage ($\varphi$), were fitted using the physical model (Methods Section 2.14b).

### 3.4 Air/water and oil/water macroscopic contact angle measurements

The mCAs were measured on the planar substrates using the standard sessile drop method (Eq. 1, Fig. 5a) with the techniques of receding (RCA) and advancing contact angle (ACA) at air/water and oil/water interfaces. As expected with the sessile drop method, large CA distributions require sufficient samples ($n = 50$) to represent the CA spread that can be expected for real systems [4,9]. The mCAs were recorded after prewetting of the substrates with larger volumes of water that were subsequently discarded, leaving behind a thin water film that more closely resembles the surface of the water-dispersed NPs. This also allows possible air-borne impurities to dissolve in the discarded liquid and it minimizes the observed stick-slip behavior on dry substrates [41]. Particularly amino-terminated functionalities are prone to quickly adsorb air- or water-borne impurities that impact the measured ACAs/RCAs, complicating the experimental execution in measuring mCAs [76]. This pretreatment decreased the air/water ACAs/RCAs compared to drops on a dry surface (data not shown) [77]. Sequential ACA/RCA cycles on the same sample immersed in decane provided similar mCAs with less pronounced stick-slip compared to the first cycle, indicated by a semi-stable ACA region that is irrespective of prewetting conditions (Supplementary Section S5, Fig. S5).

Table 2 summarizes the mean CAs and hysteresis percentages and exact numbers of all CAs for substrates and particles are given in Supplementary Section S6, Table S6. At first glance, the wettability enhances for the particle-covered surfaces compared to the smooth surfaces, which agrees to Wenzel's observation under the assumption that the liquid completely penetrates the surface grooves [78]. The RCAs at air/water for all samples are practically zero, also due to their hydrophilicity. The mean air/water ACA of smooth silica is 9.42° and the oil/water ACA/RCA is 14.12°/8.57°, while the substrate covered with silica particles shows significant spreading in both environments. The ACA of smooth APTES substrates is 29.02° and decreases to 9.61° on the particle-covered APTES surface. When decane is deposited on top of the smooth substrate, the average oil/water ACA on SF-APTES-SiO$_2$ increases to 57.01° and the RCA is 21.44°. The oil/water ACA of the particle-covered APTES film showed only a minor increase to 12.19° whereas no RCAs could be measured due to the spreading of the droplet during liquid retraction. Table 2 also shows the equilibrium contact angle (ECA, Eq. 4), which is the angle defined by the Young-Laplace equation from the far-field equilibrium of the surface forces acting on the triple line, where colloidal interactions are irrelevant, down to length scales of ~ 30 nm. Below this size, the microscopic structure of the contact line is subject to physicochemical properties of the solid and fluids along with van der Waals and electrostatic interactions exerting additional influence on the local contact angle [79]. Except for SF-APTES in decane/water, the ECA is very hydrophilic (< 18°). The contact angle hysteresis (CAH, Eq. 7) is given as a percentage to easily compare the different surfaces with each other [40]. The roughness-induced wetting for the particle-covered surfaces and smooth silica show very low hysteresis ($\Delta\theta_\% < 1\%$), while SF-APTES-SiO$_2$ shows hysteresis most likely due to topographical (Fig. 2) and chemical [39] irregularities or surface defects. The large increase in CAH for smooth APTES when decane is added is proportional to the increase in the work of spreading (Supplementary Section S7) [80]. Table 3 shows the average apparent surface free energies for the smooth surfaces to assess the solid-oil-water interactions (Eq. 12) derived from air/water ACA/RCAs and ECAs. Depending on the applied model, at least 2 test liquids are usually necessary to find the solid-gas interfacial tension ($\gamma_{sg}$) to establish the solid-gas-liquid interactions that describe the adhesion between the solid and liquid. Contrarily, the contact angle hysteresis (Eq. 8) [43] or equation of state approaches (EOS, Eq. 9) [41] roughly estimate $\gamma_{sg}$ with only one test liquid [81]. While more liquids in Eq. 8 – 9 improve accuracy in estimating $\gamma_{sg}$, similar results to ours are found for silica with extended models (e.g. Wu [82]) that employed a significant number of solvents [83]. The solid-liquid interfacial tension ($\gamma_{sl}$) determined from the EOS (Eq. 10) is used to estimate the solid-oil interfacial tension ($\gamma_{so}$), which is an extension of the EOS (Eq. 11) applicable to a solid surrounded by two immiscible fluids. Similarly, the Young equation (Eq. 1) was adapted for oil/water systems into Eq. 12 leading to an average oil/water CA of 12.64° for silica and 37.68° for APTES-coated silica when using the ACA/RCA numbers, which are close to the oil/water ECAs as determined from Eq. 4 [44]. The predicted oil/water/solid CAs determined from the air/water ECAs lie between 5.70° to 10.49° for silica and 14.75° to 24.67° for APTES-coated silica, both regions where the experimental oil/water RCAs fall into. Based on these results, air/water ACA/RCA crudely estimates the oil/water ECA, while air/water ECAs predict the oil/water RCA.

**Table 2** Advancing (ACA), receding (RCA), equilibrium contact angle (ECA, Eq. 4), and contact angle hysteresis (CAH, Eq. 7) for the smooth (SF) and particle-covered (PF) surfaces with standard deviation ($n = 50$).

| | | SF-SiO$_2$ | SF-APTES-SiO$_2$ | PF- SiO$_2$ | PF-APTES-SiO$_2$ |
|---|---|---|---|---|---|
| Air/water | **ACA** (°) | 9.42 ± 1.34 | 29.02 ± 3.48 | 0.10* - 5.00** | 9.61 ± 2.35 |
| | **RCA** (°) | 0.10* - 5.00** | 0.10* - 5.00** | 0.10* - 5.00** | 0.10* - 5.00** |
| | **ECA** (°) | 3.99 ± 0.50* - 7.31 ± 0.72** | 10.35 ± 1.02* - 17.62 ± 1.75** | 0.10* - 5.00** | 3.46 ± 0.86 - 7.42 ± 1.26 |
| | **CAH** (%) | 0.67 ± 0.19 | 6.28 ± 1.47 | 0.00 | 0.70 ± 0.35 |
| Decane/water | **ACA** (°) | 14.12 ± 1.55 | 57.01 ± 9.49 | 0.10* - 5.00** | 12.19 ± 2.50 |
| | **RCA** (°) | 8.57 ± 1.69 | 21.44 ± 6.71 | 0.10* - 5.00** | 0.10* - 5.00** |
| | **ECA** (°) | 11.45 ± 1.60 | 39.31 ± 7.67 | 0.10* - 5.00** | 4.98 ± 0.88 - 8.81 ± 1.35 |
| | **CAH** (%) | 0.95 ± 0.11 | 19.32 ± 4.78 | 0.00 | 1.13 ± 0.46 |

Due spreading of the drops, a CA of 0.10°* or 5.00°** was chosen to determine the ECA.

**Table 3** Surface free energies (mJ m$^{-2}$) for the smooth silica (SF-SiO$_2$) and APTES-coated silica (SF-APTES-SiO$_2$) substrates based on water contact angle measurements (Eq. 8 – 11) to predict the contact angle (°) for solid-water-oil (Eq. 12, $\theta_{lo}$) with standard deviations ($n = 50$).

| | | $\gamma_{sg}^{CAH} - \gamma_{sg}^{EOS}$ | $\gamma_{sg}^{Av}$ | $\gamma_{sl}$ | $\gamma_{so}$ | $\theta_{lo}$ |
|---|---|---|---|---|---|---|
| ACA and RCA | SF-SiO$_2$ | 71.46 ± 0.21 – 71.23 ± 0.27 | 71.35 ± 0.25 | 1.67 x 10$^{-2}$ ± 0.95 x 10$^{-2}$ | 50.93 ± 0.36 | 12.64 ± 1.84 |
| | SF-APTES-SiO$_2$ | 65.42 ± 1.56 – 64.36 ± 1.61 | 64.89 ± 1.53 | 1.14 ± 0.46 | 42.37 ± 1.86 | 37.68 ± 4.17 |

| | | $\gamma_{sg}^{EOS}$ | $\gamma_{sl}$ | $\gamma_{so}$ | $\theta_{lo}$ |
|---|---|---|---|---|---|
| ECA (RCA 0.1°) | SF-SiO$_2$ | 72.03 ± 0.05 | 6.75 x 10$^{-3}$ ± 3.40 x 10$^{-3}$ | 51.94 ± 0.07 | 5.70 ± 0.75 |
| | SF-APTES-SiO$_2$ | 71.05 ± 0.22 | 2.89 x 10$^{-2}$ ± 1.06 x 10$^{-2}$ | 50.50 ± 0.31 | 14.75 ± 1.39 |

| | | $\gamma_{sg}^{EOS}$ | $\gamma_{sl}$ | $\gamma_{so}$ | $\theta_{lo}$ |
|---|---|---|---|---|---|
| ECA (RCA 5°) | SF-SiO$_2$ | 71.62 ± 0.12 | 7.49 x 10$^{-3}$ ± 2.89 x 10$^{-3}$ | 51.33 ± 0.17 | 10.49 ± 1.04 |
| | SF-APTES-SiO$_2$ | 69.01 ± 0.60 | 0.22 ± 0.08 | 47.63 ± 0.81 | 24.67 ± 2.34 |

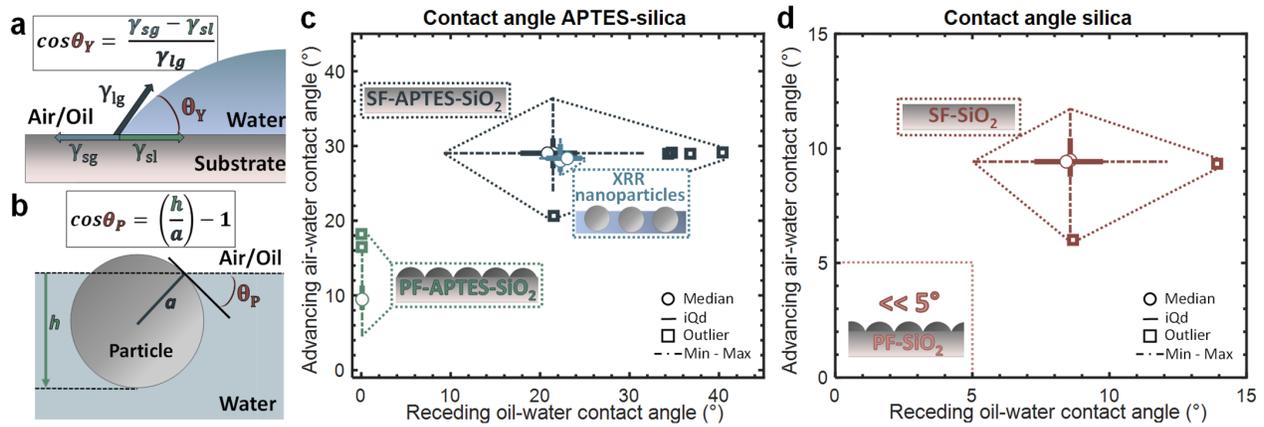

**Fig. 5 Contact angles of adsorbed nanoparticles and sessile droplets**. **a**, The standard Young equation (Eq.1) and **b**, relating immersion depth (Eq. 3) to estimate the contact angle ($\theta$) at the air/water and oil/water interface is applied to macroscopic sessile droplets on planar substrates and interfacially adsorbed nanoparticles resolved with X-ray reflectometry (XRR), respectively. **b,c** A double boxplot of the advancing air/water contact angle against the receding oil/water contact angle for APTES-coated silica (**c**) and bare silica (**d**) visualizes macroscopic contact angles measured on the smooth (SF) and particle-covered films (PF) along with the interfacially adsorbed nanoparticles obtained from XRR (**c**). Circles are the median, 'iQd' stands for interquartile distance (middle 50 % of the data), the intersection points of air/water and decane/water correspond to the mean and the dash-dotted line shows the (nonoutlier) minimum and maximum values ($n = 50$).

## 3.5 Comparing contact angles at the nano- and macroscale

The effects of gravity and long-range surface interactions (van der Waals, electrostatic, hydrophobic, and hydration forces) define the theoretical dimensions to properly relate the observed CAs for liquid droplets on planar surfaces and solid particles adsorbing at a liquid interface. The former is expressed as the capillary length (($l_c = \sqrt{\gamma/\Delta\rho g}$), where $\gamma$ is the surface/interfacial tension, $\Delta\rho$ the liquid density difference and $g$ the gravitational acceleration), which is around 2.71 mm in air/water and 4.44 mm for decane/water, while the latter operates below a characteristic length scale $l_s = \theta_E^2 \sqrt{H/6\pi\gamma} \approx$ 1-10 nm, $\theta_E$ is the ECA, and $H$ is the Hamaker constant. If the principal radii are smaller than $l_c$, gravitational forces do not influence the shape of the droplet or interface for sessile droplets or adsorbing colloids, respectively [84]. The sessile drops in this study on the substrates and the NP size are well below $l_c$ and above $l_s$. While these conditions apply to perfectly smooth and defect-free surfaces, we assume that, beyond the nanometer-scale roughness of the PF substrates (~ 35 nm), the sub-nanometer roughness on the surface between particles and smooth planar thin films are similar. Deviations from perfectly smooth surfaces give rise to greater contact line pinning both at the nano- and macroscale (and thus hysteresis) with additional line tension contributions below a droplet or particle size of 100 nm with size-dependent CAs [16]. Line tension is primarily discussed for uncharged fluids, compared to charged solids, with a strong disparity in sign and magnitude [7,8]. Our current approach assumes that equivalent $\zeta$ potentials between the NPs and planar substrates improve the predictability of particle wetting as inferred from mCAs of equivalent macroscopic substrates without exact knowledge about pinning or line tension.

It has been proposed that hydrophilic NPs adsorbing from the aqueous phase have CAs that are equivalent to the RCA, while the ACA corresponds to a particle adsorbing from the air/oil phase [84,85]. Considering the motion of an adsorbing particle towards the interface from either bulk phase, frequent and large surface defects increase the amount of contact line pinning on the particle surface as it protrudes the interface. An advancing particle adsorbing from air/oil to the liquid interface achieves a stable angle after displacement of the contact line, thereby increasing the wetted area of the colloid (higher CA compared to ECA), whereas a lowering of this wetted area occurs for the particle receding the interface (lower CA compared to ECA) [84]. One expects that hydrophilic NPs adsorbing at the interface have CAs that more closely correspond to the RCA measured on a flat substrate characterizing similar surface defects, which can be roughly estimated from the CAH. The mCAs, which encompass the ACA, RCA, and ECA, are compared to the CAs of self-assembled NPs at the liquid surface derived from XRR (Supplementary Section S6, Table S6) and calculated using Eq. 3 (Fig. 5b). Based on the XRR results, only APTES-coated silica NPs adsorb at both liquid interfaces, providing the foundation to compare the NP CAs to the respective mCAs. These corresponding results are summarized as a double boxplot in Fig. 5c and 5d for APTES-coated silica and bare silica, respectively, where the mean values lie on the intersection points of each sample.

### 3.5.1 APTES-coated silica

The obtained mean and median mCAs (air/water ACAs and decane/water RCAs) of smooth APTES-coated planar substrates, comprising similar surface chemistry and $\zeta$ potentials as the interfacially adsorbed NPs, coincide nicely with the CA distribution of the NPs measured with XRR, while the rough, particle-covered surfaces are much more hydrophilic and are unsuitable for judging the CA of adsorbed NPs (Fig. 5c). First of all, the oil/water RCA on the flat substrates describes very well the oil/water CAs of the NPs. Based on the air/water ECA and ACA/RCA data of the smooth substrates, the estimated oil/water CAs in Table 3 provide an angle of 14.75° and 37.68°, respectively. Surprisingly, we witness that the ACA at air/water, rather than the expected RCA values, complement the wetting of the NPs. In the literature, the CAs of particles adsorbed at air/water versus oil/water mostly portray an increase in CA for the same particles at the liquid/liquid system, meaning that the choice for estimating wetting at both liquid interfaces stems from either the ACA or the RCA and not a combination of them [86,87]. However, this trend is not observed for 1 μm APTES-coated silica, with an air/water CA of ~ 62° and ~ 39° at the oil/water interface [88]. The exact reason for this prominent difference is unknown, but its origin can be related to the charge sign since opposite behavior in immersion depth is observed for negatively and positively charged colloids when salt is added to the aqueous phase [89] or simply due to line tension effects as its sign reverses [8,15,90]. Experimentally, the air/water mCAs measured here and elsewhere [22,91] do not support the fact that the air/water RCA measured on a planar substrate agrees with the CA of the particles, while mCA simulations of ultrasmall NPs at air/water [92] and oil/water [93] do agree with the RCA. Further factors need to be considered, such as the influence of prewetting (which lowers the mCAs), the wettability of the particle material itself (hydrophilic/hydrophobic), and the differences between contact line formation on the macro- and nanoscale (i.e. pinning and line tension effects).

Based on this discussion, we propose a tentative adsorption mechanism for positively charged, hydrophilic NPs at the oil/water and air/water interface at a pH ~ 5.8 and absence of electrolyte (Fig. 6a & b). Before adsorption, the positively charged hydrophilic NPs are attracted towards the negative charges present at the phase boundary (Ia & Ib) [25,73,74]. Their preferred state after protrusion resembles the RCA as a consequence of particle motion and contact line pinning, experiencing a concave meniscus for the nonpolar phase (IIIa). At air/water, no particles should populate the interface according to the measured RCA (IIb), but in this case, the ACA suggests that the meniscus is convex (IVb) and that the air/water surface wets the particle. To the best of our knowledge, the origin or possibility of such a mechanism has not been explored. Another possibility that occurs only at an air/water surface is enhanced evaporation at the three-phase contact line, causing pinning/depinning effects typically observed for drops on planar substrates exhibiting stick-slip [14] and potentially cause a change in meniscus profile for adsorbed particles. Recently, a fresh insight into the origin of the charge at air/water and oil/water interfaces indicated that both carry a net negative charge of similar magnitude but basically, no charges are present in the air while a large negative charge travels through the oil due to charge transfer of the water, making the latter subsurface region more positive [74]. The possibility of varying effective charges in the vicinity of the air/water and oil/water interfaces might assist in explaining the observed difference in CAs for the same particles [86–88].

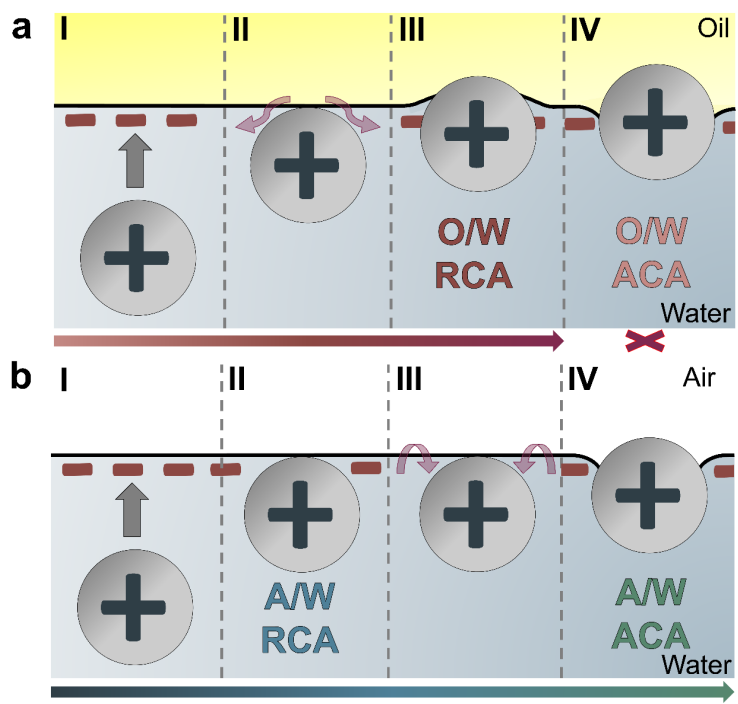

**Fig. 6 Tentative scheme relating advancing (ACA) and receding contact angle (RCA) of the planar substrates to the adsorbed APTES-coated silica nanoparticles. I**, For both oil/water (O/W, **a**) and air/water (A/W, **b**), a positively charged nanoparticle is attracted to the interface due to the net negative charges present at the phase boundary. ACAs and RCAs for adsorbing particles create oppositely oriented menisci due to contact line pinning stemming from morphological/chemical imperfections, dragging along part of the liquid. **IIa**, In the case of the O/W interface, the particle and interface interact before breaching through and pinning causes a locally deformed convex meniscus with respect to the water phase, agreeing well with the O/W RCA on smooth planar substrates (**IIIa**). **IVa**, Very high O/W ACAs are energetically unfavorable for these hydrophilic NPs and are not observed. **IIb**, Particle adsorption is not directly observed at the A/W interface, where the A/W RCA is practically zero, while the A/W ACA does correspond to the adsorbed NPs (**IIIb**). This configuration might be possible when the liquid surface wets the particle surface and is attracted towards it. The arrows positioned at the interface indicate how the meniscus pins to the particle surface, in accordance with ACA and RCA, but it remains unclear if such a change in the meniscus is physically possible when the nonpolar phase changes between gas and fluid.

### 3.5.2 Silica

Since adsorption of the bare silica particles could not be detected with XRR, a direct comparison of the CAs of particles and films cannot be performed, although some studies reported adsorption of these hydrophilic particles at the liquid surface [91,94]. The monotonic decreasing reflectivity curves for silica in Fig. 4 cannot support these observations since a previous study using XRR and grazing-incidence small-angle X-ray scattering (GISAXS) noticed that only subsurface ordering takes place where the particles just touch the air/water surface and this for very concentrated silica dispersions (40 w/w%) [95]. As no particle adsorption is detected, the measured mCAs on the smooth silica substrates in Fig 5d might represent the inherent wettability of the particle surface since the $\zeta$ potentials of the particles and substrates coincide in this case as well. When adsorption of these charged, hydrophilic silica particles is reported, external factors likely play a role such as perturbations of the liquid surface, contaminations present in the solvents, or adsorbed on the particle surface, or use of spreading solvents [21]. Supplementary Section S8, Fig. S8 shows a schematic overview of the proposed methodology when a general estimate on particle wettability is required when more sophisticated *in situ* methods are not readily available.

# 4. Conclusion

Assessing the wetting properties of small nanoparticles (NPs) necessitates resource-intensive and sophisticated approaches like synchrotron-based experiments [4,24,25,95] or other alternative methods [19–23]. Direct estimation of NP immersion depth at air/water and decane/water interfaces collected from synchrotron-based X-ray reflectivity shows that negatively charged silica ($SiO_2$) NPs do not adsorb, contrarily to the positively charged APTES-coated silica (APTES-$SiO_2$) NPs. Since wettability of particles [28] and planar substrates [29] have been linked to their surface charge, coupling of the pH-dependent $\zeta$ potential between chemically and topographically equivalent spherical particles and planar substrates ensures higher accuracy than sessile drop-derived macroscopic contact angles (mCAs) alone. Smooth or particle-covered films with similar coatings as the dispersed particles were prepared to account for possible surface roughness effects. The planar substrates measured with streaming current showed similar $\zeta$ potentials as the bulk particle dispersions analyzed through electrophoresis and were almost insensitive to surface roughness. Contrarily, the particle-covered substrates with increased surface roughness showed very hydrophilic behavior determined from the mCAs. In the case of the smooth APTES-coated surface, higher air/water advancing (ACAs) and lower oil/water receding contact angles (RCAs) provided the most realistic description of the NP wetting properties, similar to the observations for microparticles [88]. However, it is generally expected that the RCA better describes wetting of particles adsorbing from the aqueous bulk to the interface and an increase in the particle CA can be expected when gas is exchanged for a denser oil [22,84,91–93,96]. Apparent surface free energies derived from air/water ACAs, RCAs, and equilibrium contact angles (ECAs) provide a range of oil/water mCAs, which conforms with the experimental oil/water RCAs measured on the smooth APTES-coated surfaces [13,41–44]. At the same time, the mCA inherently can only provide approximations of the CAs of straddling particles at a liquid surface. Additionally, the high accuracy of X-ray scattering in this context opens up new possibilities to experimentally investigate line tension contributions of charged particle surfaces [8] and CA dependencies by varying particle size, pH, electrolyte concentration in either phase and/or using immiscible oils with different polarities, shedding more light on the applicability of ACAs/RCAs with a variety of particle materials and environmental conditions.

# CRediT authorship contribution statement

**Joeri Smits:** Data Curation, Formal analysis, Investigation, Methodology, Validation, Visualization, Writing – Original Draft, Writing – Review & Editing. **Rajendra Prasad Giri:** Data Curation, Formal analysis, Investigation, Validation, Visualization, Writing – Review & Editing. **Chen Shen:** Data Curation, Formal analysis, Investigation, Validation, Visualization, Writing – Review & Editing. **Diogo Mendonça:** Data Curation, Formal analysis, Investigation, Validation, Visualization, Writing – Review & Editing. **Bridget Murphy:** Data Curation, Investigation, Resources, Supervision, Validation, Visualization, Writing – Review & Editing. **Patrick Huber:** Data Curation, Investigation, Resources, Supervision, Validation, Visualization, Writing – Review & Editing. **Kurosch Rezwan:** Resources, Supervision, Validation, Writing – Review & Editing. **Michael Maas:** Conceptualization, Data Curation, Investigation, Funding acquisition, Methodology, Project administration, Resources, Supervision, Validation, Visualization, Writing – Review & Editing.

# Declaration of Competing Interest

The authors declare that they have no known competing financial interests or personal relationships that could have appeared to influence the work reported in this paper.

# Acknowledgements

The authors thank P. Witte for her expertise with the SEM and T. Mehrtens for TEM. They acknowledge DESY (Hamburg, Germany), a member of the Helmholtz Association HGF, for the provision of experimental facilities. Parts of this research were carried out at PETRA III, and the authors thank Dr. M. Lippmann for assistance in using the lab facility. The authors acknowledge research grant Verbundforschung BMBF/05KS7FK3/05KS10FK2, Verbundforschung

BMBF/05K16FK1/05K19FK2 and 05K19K2 for financing LISA instrument and the Lambda GaAs detector. M.M. acknowledges support by DFG (Deutsche Forschungsgemeinschaft) project 278836263. P.H. acknowledges support by the DFG Graduate School GRK2462 "Processes in natural and technical Particle-Fluid-Systems (PintPFS)" (Project number 390794421). This work was also supported by the DFG within the Collaborative Research Initiative SFB 986 "Tailor-Made Multi-Scale Materials Systems" (project number 192346071).

# Figure captions

**Fig. 1 Overview of the methodology to relate the particle contact angle at liquid surfaces to macroscopic contact angles on planar substrates.** The particles and planar substrates consist of silica ($SiO_2$) or are coated with APTES (APTES-$SiO_2$). The liquid interface in presence of 18 nm $SiO_2$ and 19 nm APTES-$SiO_2$ is analyzed with synchrotron X-ray reflectometry from which the particle contact angle is derived. The zeta potential of particles and planar substrates is measured and compared using electrophoresis and streaming current, respectively. The effect of surface roughness is investigated by comparing the wettability of a smooth surface (smooth film, SF) with a particle-induced rough surface (particle-covered film, PF).

**Fig. 2 Topographical analysis of planar smooth (SF) and particle-covered films (PF) of silica ($SiO_2$) and APTES-coated silica (APTES-$SiO_2$) with AFM and SEM.** A 3D surface area of 10 x 10 μm obtained with AFM is shown from which the average root mean square (RMS) roughness per type is determined with standard deviation ($n = 4$). The bottom row depicts SEM images with scale bars of 1 μm.

**Fig. 3 Evaluation of zeta potential with electrophoresis and streaming current at 10 mM KCl**. **a**, Zeta potential measurements using electrophoretic light scattering (EP) for 0.1 w/w% of 1000, 80, and 18 nm silica ($SiO_2$) particles at different pH with their approximate dimensionless electrokinetic radius $\varkappa a$. **b**, Zeta potential using the streaming current (SC) method for smooth (SF) and particle-covered films (PF) with $SiO_2$ or coated with APTES (APTES-$SiO_2$) are measured in a pH range from 5.8 to 9.4. **c**, Comparing the zeta potentials obtained with the ELS and SC method of $SiO_2$ and APTES-$SiO_2$. Only results of the SF-samples from **3b** are depicted in **c**. The maximum range of the pH titration was different for both methods and is indicated schematically by the broad arrows and experimental standard deviations are provided in all plots ($n = 3$).

**Fig. 4 X-ray reflectivity and electron densities of adsorbed particles at the air/water surface and decane/water interface. a**, Reflectivity curves normalized to the Fresnel reflectivity ($R_F$) at the air/water interface and **b**, the decane/water interface (aqueous subphase pH ∼ 5.8) with 0.1 w/w% 18 nm silica ($SiO_2$, left triangles) and 19 nm APTES-coated silica (APTES-$SiO_2$, right triangles) with their corresponding fits (black line) and standard deviation ($n = 3$). The insets show the normalized-to-water ($\rho_w$) electron density plots for $SiO_2$ (dashed line) and APTES-$SiO_2$ (solid line) from which the physical parameters, such as particle coverage ($\varphi$), were fitted using the physical model (Methods Section 2.14b).

**Fig. 5 Contact angles of adsorbed nanoparticles and sessile droplets**. **a**, The standard Young equation (Eq.1) and **b**, relating immersion depth (Eq. 3) to estimate the contact angle ($\theta$) at the air/water and oil/water interface is applied to macroscopic sessile droplets on planar substrates and interfacially adsorbed nanoparticles resolved with X-ray reflectivity (XRR), respectively. **b,c** A double boxplot of the advancing air/water contact angle against the receding oil/water contact angle

for APTES-coated silica (**c**) and bare silica (**d**) visualizes macroscopic contact angles measured on the smooth (SF) and particle-covered films (PF) along with the interfacially adsorbed nanoparticles obtained from XRR (**c**). Circles are the median, 'iQd' stands for interquartile distance (middle 50 % of the data), the intersection points of air/water and decane/water correspond to the mean and the dash-dotted line shows the (nonoutlier) minimum and maximum values ($n$ = 50).

**Fig. 6 Tentative scheme relating advancing (ACA) and receding contact angle (RCA) of the planar substrates to the adsorbed APTES-coated silica nanoparticles. I**, For both oil/water (O/W, **a**) and air/water (A/W, **b**), a positively charged nanoparticle is attracted to the interface due to the net negative charges present at the phase boundary. ACAs and RCAs for adsorbing particles create oppositely oriented menisci due to contact line pinning stemming from morphological/chemical imperfections, dragging along part of the liquid. **IIa**, In the case of the O/W interface, the particle and interface interact before breaching through and pinning causes a locally deformed convex meniscus with respect to the water phase, agreeing well with the O/W RCA on smooth planar substrates (**IIIa**). **IVa**, Very high O/W ACAs are energetically unfavorable for these hydrophilic NPs and are not observed. **IIb**, Particle adsorption is not directly observed at the A/W interface, where the A/W RCA is practically zero, while the A/W ACA does correspond to the adsorbed NPs (**IIIb**). This configuration might be possible when the liquid surface wets the particle surface and is attracted towards it. The arrows positioned at the interface indicate how the meniscus pins to the particle surface, in accordance with ACA and RCA, but it remains unclear if such a change in the meniscus is physically possible when the nonpolar phase changes between gas and fluid.

# Tables

**Table 1** Overview of the used silica (SiO$_2$) and APTES-coated silica (APTES-SiO$_2$) materials and their main purpose. ACA, advancing contact angle; AFM, atomic force microscope; ELS, electrophoretic light scattering; mCA, macroscopic contact angle; RCA, receding contact angle; RMS, root mean square; XRR, X-ray reflectivity.

| Material | Size | Main purpose |
|---|---|---|
| SiO$_2$ | 17.8 nm | ELS: $\zeta$ potential; XRR: air/water and decane/water interfacial adsorption and CA |
| | 80 nm | ELS: $\zeta$ potential |
| | 1000 nm | ELS: $\zeta$ potential |
| | Planar (smooth and particle-covered) | AFM: RMS roughness; mCA: air/water and decane/water ACA/RCA; SC: $\zeta$ potential |
| APTES-SiO$_2$ | 18.9 nm | ELS: $\zeta$ potential; XRR: air/water and decane/water interfacial adsorption and CA |
| | Planar (smooth and particle-covered) | AFM: RMS roughness; mCA: air/water and decane/water ACA/RCA; SC: $\zeta$ potential |

**Table 2** Advancing (ACA), receding (RCA), equilibrium contact angle (ECA, Eq. 4), and contact angle hysteresis (CAH, Eq. 7) for the smooth (SF) and particle-covered (PF) surfaces with standard deviation ($n = 50$).

| | | SF-SiO$_2$ | SF-APTES-SiO$_2$ | PF- SiO$_2$ | PF-APTES-SiO$_2$ |
|---|---|---|---|---|---|
| Air/water | **ACA** (°) | 9.42 ± 1.34 | 29.02 ± 3.48 | 0.10* - 5.00** | 9.61 ± 2.35 |
| | **RCA** (°) | 0.10* - 5.00** | 0.10* - 5.00** | 0.10* - 5.00** | 0.10* - 5.00** |
| | **ECA** (°) | 3.99 ± 0.50* - 7.31 ± 0.72** | 10.35 ± 1.02* - 17.62 ± 1.75** | 0.10* - 5.00** | 3.46 ± 0.86 - 7.42 ± 1.26 |
| | **CAH** (%) | 0.67 ± 0.19 | 6.28 ± 1.47 | 0.00 | 0.70 ± 0.35 |
| Decane/water | **ACA** (°) | 14.12 ± 1.55 | 57.01 ± 9.49 | 0.10* - 5.00** | 12.19 ± 2.50 |
| | **RCA** (°) | 8.57 ± 1.69 | 21.44 ± 6.71 | 0.10* - 5.00** | 0.10* - 5.00** |
| | **ECA** (°) | 11.45 ± 1.60 | 39.31 ± 7.67 | 0.10* - 5.00** | 4.98 ± 0.88 - 8.81 ± 1.35 |
| | **CAH** (%) | 0.95 ± 0.11 | 19.32 ± 4.78 | 0.00 | 1.13 ± 0.46 |

Due spreading of the drops, a CA of 0.10°* or 5.00°** was chosen to determine the ECA.

**Table 3** Surface free energies (mJ m$^{-2}$) for the smooth silica (SF-SiO$_2$) and APTES-coated silica (SF-APTES-SiO$_2$) substrates based on water contact angle measurements (Eq. 8 – 11) to predict the contact angle (°) for solid-water-oil (Eq. 12, $\theta_{lo}$) with standard deviations ($n = 50$).

| | | $\gamma_{sg}^{CAH}$ - $\gamma_{sg}^{EOS}$ | $\gamma_{sg}^{Av}$ | $\gamma_{sl}$ | $\gamma_{so}$ | $\theta_{lo}$ |
|---|---|---|---|---|---|---|
| ACA and RCA | SF-SiO$_2$ | 71.46 ± 0.21 – 71.23 ± 0.27 | 71.35 ± 0.25 | 1.67 x 10$^{-2}$ ± 0.95 x 10$^{-2}$ | 50.93 ± 0.36 | 12.64 ± 1.84 |
| | SF-APTES-SiO$_2$ | 65.42 ± 1.56 – 64.36 ± 1.61 | 64.89 ± 1.53 | 1.14 ± 0.46 | 42.37 ± 1.86 | 37.68 ± 4.17 |

| | | $\gamma_{sg}^{EOS}$ | $\gamma_{sl}$ | $\gamma_{so}$ | $\theta_{lo}$ |
|---|---|---|---|---|---|
| ECA (RCA 0.1°) | SF-SiO$_2$ | 72.03 ± 0.05 | 6.75 x 10$^{-3}$ ± 3.40 x 10$^{-3}$ | 51.94 ± 0.07 | 5.70 ± 0.75 |
| | SF-APTES-SiO$_2$ | 71.05 ± 0.22 | 2.89 x 10$^{-2}$ ± 1.06 x 10$^{-2}$ | 50.50 ± 0.31 | 14.75 ± 1.39 |

| | | $\gamma_{sg}^{EOS}$ | $\gamma_{sl}$ | $\gamma_{so}$ | $\theta_{lo}$ |
|---|---|---|---|---|---|
| ECA (RCA 5°) | SF-SiO$_2$ | 71.62 ± 0.12 | 7.49 x 10$^{-3}$ ± 2.89 x 10$^{-3}$ | 51.33 ± 0.17 | 10.49 ± 1.04 |
| | SF-APTES-SiO$_2$ | 69.01 ± 0.60 | 0.22 ± 0.08 | 47.63 ± 0.81 | 24.67 ± 2.34 |